\documentclass[twocolumn,english,aps,prb, superscriptaddress]{revtex4}

\usepackage[T1]{fontenc}
\usepackage[utf8]{inputenc}
\usepackage{amsmath}
\usepackage{amssymb}
\usepackage{graphicx}

\makeatletter
\@ifundefined{textcolor}{}
{%
 \definecolor{BLACK}{gray}{0}
 \definecolor{WHITE}{gray}{1}
 \definecolor{RED}{rgb}{1,0,0}
 \definecolor{GREEN}{rgb}{0,1,0}
 \definecolor{BLUE}{rgb}{0,0,1}
 \definecolor{CYAN}{cmyk}{1,0,0,0}
 \definecolor{MAGENTA}{cmyk}{0,1,0,0}
 \definecolor{YELLOW}{cmyk}{0,0,1,0}
}

\usepackage[caption=false]{subfig}
\usepackage{bm}\usepackage{amsfonts}\usepackage{dsfont}\@ifundefined{definecolor}
 {\usepackage{color}}{}
\usepackage{enumerate}

\makeatother

\usepackage{babel}
\begin{document}

\title{Topological phases in gapped edges of fractionalized systems}
\begin{abstract}
Recently, it has been proposed that exotic one-dimensional phases can be realized by gapping out the edge states of a fractional topological insulator. The low-energy edge degrees of freedom are described by a chain of coupled parafermions. We introduce a classification scheme for the phases that can occur in parafermionic chains. We find that the parafermions support both topological symmetry fractionalized phases as well as 
phases in which the parafermions condense. 
In the presence of additional symmetries, the  phases form a non-Abelian group.  As a concrete example of the classification, we consider the effective edge model for a $\nu= 1/3$ fractional topological insulator for which we calculate the entanglement spectra numerically and show that all possible predicted phases can be realized.
\end{abstract}

\author{Johannes Motruk}

\affiliation{Max-Planck-Institut für Physik komplexer Systeme, Nöthnitzer Strasse
38, 01187 Dresden, Germany}

\author{Erez Berg}

\affiliation{Department of Condensed Matter Physics, Weizmann Institute of Science, Rehovot, Israel 76100}

\author{Ari M. Turner}
\affiliation{Institute for Theoretical Physics, University of Amsterdam, Science Park 904, P.O. Box 94485, 1090 GL Amsterdam, The Netherlands}

\author{Frank Pollmann}

\affiliation{Max-Planck-Institut für Physik komplexer Systeme, Nöthnitzer Strasse
38, 01187 Dresden, Germany}

\maketitle
One-dimensional quantum systems can support topologically nontrivial
phases.\cite{Berg-2008,Gu-2009,Pollmann-2010,Fidkowski-2011,Turner-2011,Pollmann12,ChenGu-2011,ChenGu-2011-2,Schuch-2011} These are gapped phases whose ground states do not break any
symmetry of the Hamiltonian, yet they cannot be smoothly connected
to a trivial (site-factorizable) state. Some of these phases also
possess fractionalized edge states, which cannot disappear unless
the bulk gap closes. In many cases, the topological phase retains
its nontrivial nature only as long as certain symmetries are kept;
a prime example of this phenomenon is the Haldane phase of integer
spin chains.\cite{Haldane-1983,Haldane-1983a} Intriguingly, paired fermions can support a phase which
is not protected by any symmetry (besides the intrinsic parity symmetry
of fermions), and is therefore robust to arbitrary perturbations.\cite{Fidkowski-2011,Turner-2011}
This is the one-dimensional topological superconductor phase, which
is characterized by fractionalized Majorana zero modes at its boundaries.\cite{Kitaev-2001} These zero modes resemble the ones found in the cores of vortices in topological superconductors in two dimensions.\cite{Read1999}
Other than the fundamental interest in this phase, it has been shown to
be potentially useful for quantum information processing. \cite{Alicea-2012} It has recently been proposed that these phases can be realized in quantum wires proximity-coupled to superconductors.\cite{Lutchyn2010, Oreg2010} Signatures of Majorana zero modes have been observed in recent experiments.\cite{Mourik-2012,Das-2012}

These breakthroughs raise the question whether there is a richer variety
of robust topological phases that can be realized in strongly-interacting
one dimensional systems. Recently, it has been argued by Turner et
al. \cite{Turner-2011}, and simultaneously by Fidkowski et al. \cite{Fidkowski-2011}
that a one-dimensional system of fermions with arbitrary interactions
and no symmetry (other than fermion parity symmetry)
can support only two phases: the trivial phase and the topological
superconductor phase.

In this paper, we discuss a possible way around this no-go argument.
One starts from an effectively one-dimensional system, which
lives on the edge of a higher-dimensional fractionalized phase (e.g.,
a fractional quantum Hall phase). Therefore, the elementary degrees
of freedom in the one-dimensional system need not be fermions or bosons;
they can be fractionalized anyonic quasiparticles of the underlying
two-dimensional phase. This leads to new classes of topological phases,
which can never be realized in a strictly one-dimensional system in the absence of symmetry.
We focus on the recently discovered example of counter-propagating
fractional quantum Hall edge states, which are gapped by proximity
coupling to a superconductor.\cite{Lindner,Clarke,Cheng} This system can be mapped to a one-dimensional
chain of particles with parafermionic statistics\cite{Fradkin-1980}, of the type discussed
recently by Fendley\cite{Fendley}. This proposal generalizes an earlier setup that consists of edges of a quantum spin Hall insulator coupled to a superconductor.\cite{Fu2008} The boundary between regions of the edge in different topological phases support new types of zero modes with non-Abelian properties.\cite{comment-2d}

In this paper we classify the possible phases of this system
on general grounds, and demonstrate the stability of the zero modes
that occur between different phases. The different phases can be distinguished
by the symmetry properties of their entanglement states, which are
reflected in characteristic degeneracies of their entanglement spectra.

This paper is organized as follows: In Sec. \ref{sec:fermion_class}
we review the arguments leading to the classification of gapped fermionic
phases in one dimension. In Sec. \ref{sec:parafermionic_chain} we
discuss a one-dimensional model of parafermions and introduce a classification scheme for parafermions in terms of the possible fractionalizations of the symmetry operator. We derive an addition rule for combining chains in different phases, and discuss how the classification is enriched when additional symmetries are present.
In Sec. \ref{sec:physical_system} we first review how to generate parafermions on the  edge of a two-dimensional (2D) fractional topological
insulator. We then introduce a setup, which allows us to realize different  topological phases on the edge of a $\nu=\frac13$ fractional topological insulator and study the effective edge model
numerically. The different phases are identified by their characteristic entanglement spectra. We summarize the main results and conclude in Sec. \ref{sec:Conclusions}.

\section{Classification of symmetry protected topological phases\label{sec:fermion_class}}

\begin{figure}
\begin{centering}
\includegraphics[width=8cm]{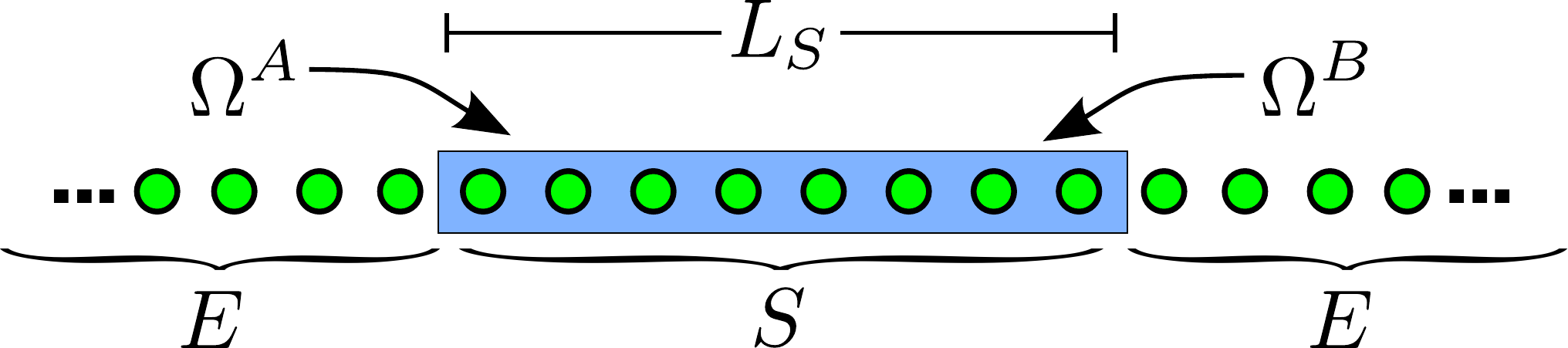}
\par\end{centering}

\caption{Illustration of a bipartition of a chain into a segment (\ensuremath{S})
of length \ensuremath{L_S} and an environment (\ensuremath{E}). A
symmetry operation $\Omega$ acting on the important eigenstates of
the reduced density matrix fractionalizes into two operators $\Omega^{A}$
and $\Omega^{B}$ .\label{fig:segment}}
\end{figure}

We begin by reviewing briefly the arguments of Ref.~\onlinecite{Turner-2011}
leading to the classification of gapped phases of interacting fermions
in one dimension. The main idea is to study the properties of the entanglement
in the ground-state wave function $|\psi_0\rangle$.  We introduce a bipartition
of the system into a segment $S$ of $L_S$ consecutive sites and an
environment $E$ (see Fig.~\ref{fig:segment}). We assume that $L_S$
is much larger than the correlation length $\xi$. We will now study
the eigenvalues $\rho_{\alpha}^{S}$ and the corresponding eigenstates
$|\phi_{\alpha}\rangle$ of the reduced density matrix $\rho^{S}=\mathrm{Tr}_E|\psi_{0}\rangle\langle\psi_{0}|$
for the ground state $|\psi_{0}\rangle$. The eigenvalues $\rho_{\alpha}^{S}$
are directly related to the so-called entanglement spectrum\cite{Li-2008} $\epsilon_{\alpha}^{S}=-\log\rho_{\alpha}^{S}$.
As the system is gapped and has short range correlations, only
a finite number of states $|\phi_{\alpha}\rangle$ have an eigenvalue
$\rho_{\alpha}^{S}>\epsilon$ for any finite $\epsilon>0$. This is directly related to the
area law. \cite{Hastings-2007} In other
words, only a finite number of eigenstates contribute significantly
to the state $|\psi_{0}\rangle$. We will refer to these states as
the important states. It was realized in Ref.~\onlinecite{Turner-2011}
that any symmetry operation $\Omega$ acting on the important states
$|\phi_{\alpha}\rangle$ can be split into two parts $\Omega\approx\Omega^{A}\Omega^{B}$,
which are only acting on the left and right side of the segment. Different
topological phases are then characterized by the algebras of the fractionalized
symmetry operators $\Omega^{A}$ and $\Omega^{B}$.

Let us consider the example of a one-dimensional spinless superconductor
with fermion parity symmetry described by the following Hamiltonian

\begin{eqnarray}
H_{0} & = & -\frac{t}{2}\sum_{j=1}^{L-1}\left(c_{j}^{\dagger}c_{j+1}^{\dagger}+c_{j}^{\dagger}c_{j+1}^{\vphantom{\dagger}}+\mbox{H.c.}\right)+u\sum_{j=1}^{L}c_{j}^{\dagger}c_{j}^{\vphantom{\dagger}}\nonumber\\
\label{eq:fermions}
\end{eqnarray}
with $t,u\geq0$. The operator $c_{j}^{\dagger}$ ($c_{j}^{}$) creates
(annihilates) a spinless fermion on site $j$. The first term comprises
hopping of fermions as well as the creation and annihilation of pairs
of fermions while the second term acts as a chemical potential. The
fermion parity operator $Q=e^{i\pi\sum_{j}n_{j}}$ (with $n_{j}=c_{j}^{\dagger}c_{j}$)
commutes with $H_{0}$ because the total number of fermions $N_{\text{total}}$
modulo two is conserved. By defining the transformations
\begin{eqnarray}
\gamma_{2j-1} & = & c_{j}+c_{j}^{\dagger}\\
\gamma_{2j} & = & -i(c_{j}-c_{j}^{\dagger}),
\end{eqnarray}
the Hamiltonian (\ref{eq:fermions}) can be mapped (up to an
additive constant) to
\begin{eqnarray}
H & = & \frac{i}{2}\left(t\sum_{j=1}^{L-1}\gamma_{2j}\gamma_{2j+1}+u\sum_{j=1}^{L}\gamma_{2j-1}\gamma_{2j}\right).
\end{eqnarray}
The operators $\gamma_{j}$ are Majorana operators, which obey the
relations $\{\gamma_{i},\gamma_{j}\}=2\delta_{ij}$, $\gamma_{i}^{\vphantom{\dagger}}=\gamma_{i}^{\dagger}$.
We can identify two topological phases, which are characterized
by $Q$ either being fractionalized into two bosonic or two fermionic
operators $Q^{A},Q^{B}$ acting on the edges of the segment, i.e.,
we find two phases with $Q^{A}Q^{B}=e^{i\mu}Q^{B}Q^{A}$ where $\mu=0,\pi$.
For any system with $\mu=\pi$, all eigenvalues $\rho_{\alpha}$ come
in degenerate pairs. To see this, note that $Q$ and the reduced density
matrix $\rho_{S}$ can be diagonalized simultaneously. If $\rho^{S}|\phi_{\alpha}\rangle=\rho_{\alpha}^{S}|\phi_{\alpha}\rangle$
and $Q|\phi_{\alpha}\rangle=q|\phi_{\alpha}\rangle$, then the state
$|\psi_{\alpha}\rangle=Q^{A}|\phi_{\alpha}\rangle$ has the same eigenvalue
for the reduced density matrix $\rho^{S}|\psi_{\alpha}\rangle=\rho_{\alpha}^{S}|\psi_{\alpha}\rangle$
but $Q|\psi_{\alpha}\rangle=-q|\psi_{\alpha}\rangle$\emph{
}and thus\emph{ }$|\psi_{\alpha}\rangle$ is an orthogonal eigenstate.
Consider now the Hamiltonian in Eq.~(\ref{eq:fermions}) with $u=0$,
$t=1$. In this case, it is rather easy to show that there are only
two eigenstates $|\phi_{\alpha}\rangle$ on the segment $S$ (with
a nonzero weight in the density matrix). The fermion parity of these
states is given by $Q=-iQ^{A}Q^{B}\text{,}$ where $Q^{A}=\gamma_{1}$
and $Q^{B}=\gamma_{2L}$ and thus $\mu=\pi$ as the two operators
anticommute. In the case $u=1,\ t=0$ the ground state is
a simple product state and we find $\mu=0$. In the presence of both parity $Q^{2}=\mathds{1}$ and time-reversal
symmetry $T^{2}=\mathds1$, eight different topological phases are
found (see Ref.~\onlinecite{Turner-2011} for details).

\section{Parafermions on a chain\label{sec:parafermionic_chain}}
In this section, we generalize the concepts introduced in the previous sections to a chain of parafermions. Such a system was recently investigated by Fendley in Ref.~\onlinecite{Fendley} where it was shown that it exhibits exact zero modes, analogous to the Majorana edge modes.

\label{subsec:para_phases}
\subsection{Parafermionic model}
We  consider a one-dimensional model of $\mathds{Z}_N$ parafermions with the following commutation relations for \ensuremath{i<j}:

\begin{equation}
\chi_{i}\chi_{j}  =
e^{2\pi i/N}\chi_{j}\chi_{i}.\label{eq:comm}
\end{equation}
Here, $\chi_i$ are parafermion operators satisfying
\begin{equation}
\chi_{i}^{N}=1 \quad \text{and} \quad \chi_i^{\dagger} = \chi_i^{N-1}.
\end{equation}

The  parafermions are coupled between neighboring sites and
we choose alternating coupling parameters \ensuremath{t} and \ensuremath{u}. The Hamiltonian for a chain of $2L$ sites with open boundary conditions reads\cite{Fendley}
\begin{multline}
H=t\sum_{j=1}^{L-1}\left(e^{\pi i/N}\chi_{2j}^{\dagger}\chi_{2j+1}^{\vphantom{\dagger}}+e^{-\pi i/N}\chi_{2j+1}^{\dagger}\chi_{2j}^{\vphantom{\dagger}}\right)\\
+u\sum_{j=1}^{L}\left(e^{\pi i/N}\chi_{2j-1}^{\dagger}\chi_{2j}^{\vphantom{\dagger}}+e^{-\pi i/N}\chi_{2j}^{\dagger}\chi_{2j-1}^{\vphantom{\dagger}}\right).\label{eq:H_para}
\end{multline}

This Hamiltonian conserves the charge modulo $N$, i.e., it commutes
with the operator 
\begin{equation}
Q=\prod_{j=1}^L\left( -e^{-\pi i/N} \chi_{2j-1}^{\dagger}\chi_{2j}^{\vphantom{\dagger}} \right)\label{charge}
\end{equation}
which measures the total $\mathds{Z}_{N}$ charge.

It is useful to map Eq.~\eqref{eq:H_para} to the $N$-state quantum clock model using  a Jordan-Wigner transformation as shown in Ref.~\onlinecite{Fendley},
\begin{eqnarray}
\chi_{2j-1}^{\vphantom{\dagger}}=\left(\prod_{k<j}\tau_{k}\right)\sigma_{j},\ \ \chi_{2j}^{\vphantom{\dagger}}= - e^{\pi i/N}\left(\prod_{k\le j}\tau_{k}\right)\sigma_{j},
\label{eq:mapping}
\end{eqnarray}
with the matrices
\begin{eqnarray}
\sigma & = & \left(\begin{array}{ccccc}
0 & 1 & 0 & \ldots & 0\\
0 & 0 & 1 & \ldots & 0\\
\vdots & \vdots & \vdots & \ddots & \vdots\\
0 & 0 & 0 & \ldots & 1\\
1 & 0 & 0 & \ldots & 0
\end{array}\right),\nonumber \\[12pt]
\tau & = & \operatorname{diag}\left(1,e^{2\pi i/N},e^{4\pi i/N},\ldots,e^{2(N-1)\pi i/N}\right).
\end{eqnarray}
Note that 
the operators $\tau_j$, $\sigma_j$ are defined on a lattice with $L$ sites. 
The $\tau_j$, $\sigma_j$ operators satisfy
\begin{eqnarray}
\sigma_j^{N} & = & \tau_j^{N}=1\\
\sigma_j^{N-1} & = & \sigma_j^{\dagger}\\
\tau_j^{N-1} & = & \tau_j^{\dagger}.
\end{eqnarray}
The commutation relation between $\sigma$, $\tau$ is given by
\begin{equation}
\sigma_i\tau_j=e^{2\pi i \delta_{i,j} /N}\tau_j\sigma_i.
\end{equation}
It can be easily verified that the operators defined in Eq.~\eqref{eq:mapping} fulfill the parafermionic algebra.
Inserting the transformed operators in Eq.~\eqref{eq:H_para} yields
directly
\begin{equation}
H=-t\sum_{j=1}^{L-1}\left(\sigma_{j}^{\dagger}\sigma_{j+1}^{\vphantom{\dagger}}+\sigma_{j+1}^{\dagger}\sigma_{j}^{\vphantom{\dagger}}\right)
-u\sum_{j=1}^{L}\left(\tau_{j}^{\vphantom{\dagger}}+\tau_{j}^{\dagger}\right)\label{eq:H_Potts},
\end{equation}
In the case of periodic boundary
conditions, one has to pay attention to a phase change in the coupling
of $\sigma_{L}$ and $\sigma_{1}$, which becomes
\begin{eqnarray}
-e^{\pi i/N}\chi_{2L}^{\dagger}\chi_{1}^{\vphantom{\dagger}}=\sigma_{L}^{\dagger}\left(\prod_{k\le L}\tau_{k}^\dagger\right)\sigma_{1}.
\end{eqnarray}

The phase diagram of the \ensuremath{N}-state quantum clock model
has already been investigated.\cite{Matsuo2006} It is in the same universality class
as the $\mathds{Z}_N$ Villain model, which was examined by Elitzur
\emph{et al.}\cite{Elitzur79}. For \ensuremath{N<5}, it exhibits
two phases. For \ensuremath{t>u}, we obtain an ordered phase with
an \ensuremath{N}-fold degenerate ground state. If \ensuremath{t<u},
we are in the paramagnetic phase and the ground state is unique.
The phases are separated by a critical point at \ensuremath{t=u}.
For \ensuremath{N\geq5}, however, this critical point is extended
and a critical phase emerges in between the two phases in a finite
parameter region. The phase transitions into that critical phase are of
Berezinskii-Kosterlitz-Thouless (BKT) type with an essential singularity in
the correlation length. The phase itself is a BKT critical phase.

\begin{figure}
\begin{centering}
\includegraphics[width=8cm]{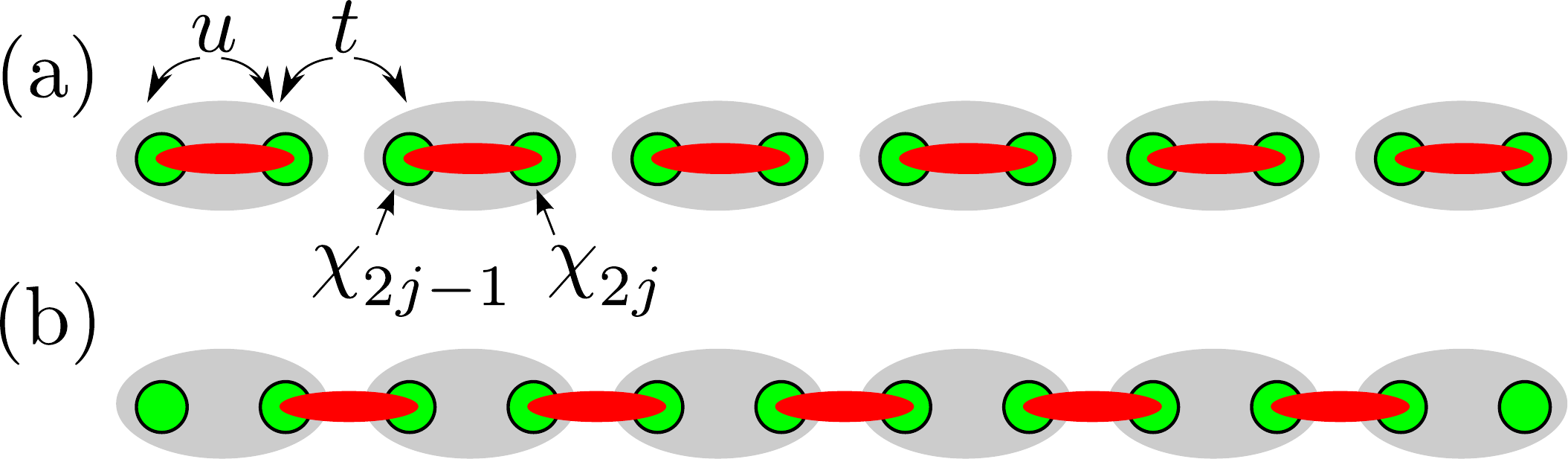}
\par\end{centering}

\caption{Two different dimerization patterns corresponding to the two phases
of the parafermionic chain. The trivial case ($u=1$, $t=0$) is
shown in (a) where pairs of para fermions $\chi_{j}$ are localized
on each site, yielding a unique and gapped ground state. Panel (b)
shows a nontrivial state ($u=0$, $t=1$) with uncoupled para fermions
at the two ends of the chain. This state is gapped in the bulk but
has gapless excitations resulting from the unpaired para fermions
and the edges. \label{fig:chain}}
\end{figure}

To understand the phase diagram of Hamiltonian Eq.~\eqref{eq:H_para},
we follow Ref.~\onlinecite{Fendley} by considering two limiting cases. The case $t=0,u=1$ [Fig.~\ref{fig:chain}(a)]
has a unique and gapped ground state, which is a factorized state where
the parafermions $\chi_{j}$ form localized pairs on each site. The
case $t=1,u=0$ [Fig.~\ref{fig:chain}(b)] is more interesting. Here,
the parafermions form pairs between neighboring sites, leaving behind
two unpaired parafermions. In the case of open boundary conditions,
the unpaired parafermions reside at the ends of the chain and yield
an $N$-fold ground-state degeneracy while the bulk remains gapped.
This can be directly compared to the case of the fermionic model discussed
in the previous section where unpaired fermionic modes appear at the
boundary. In this limit, the individual terms of $H$ commute with
each other and can be diagonalized simultaneously. Any ground state
$\left|\psi_{0}\right\rangle $ satisfies $-e^{\pi i/N}\chi_{2j}^{\dagger}\chi_{2j+1}^{\vphantom{\dagger}}\left|\psi_{0}\right\rangle =\left|\psi_{0}\right\rangle $.
This implies that projection of the charge operator $Q$ in  Eq.~(\ref{charge})
onto the ground-state manifold can be written as  $Q_{\mathrm{eff}} \propto \chi_1^\dagger \chi_{2L}^{\vphantom{\dagger}}$.
In this sense, one can say that in the low-energy manifold, the charge operator ``fractionalizes'' into a product of two operators localized at either end of the chain. Since $\chi_{1}$ and $\chi_{2L}$ both commute with the Hamiltonian, but do not commute with each other, the ground state is multiply degenerate. One can diagonalize $H$ and $Q$ simultaneously, in which case there are $N$ orthogonal ground states with distinct $Q$ eigenvalues. Acting with either $\chi_{1}^{\dagger}$
or $\chi_{2L}^{\vphantom{\dagger}}$ transforms the ground states into another, as can be seen from the fact that $\chi_1^{\dagger} Q = e^{-2\pi i/N} Q \chi_1^{\dagger}$ and similarly for $\chi^{\vphantom{\dagger}}_{2L}$.

\begin{figure}
\begin{centering}
\includegraphics[width=8cm]{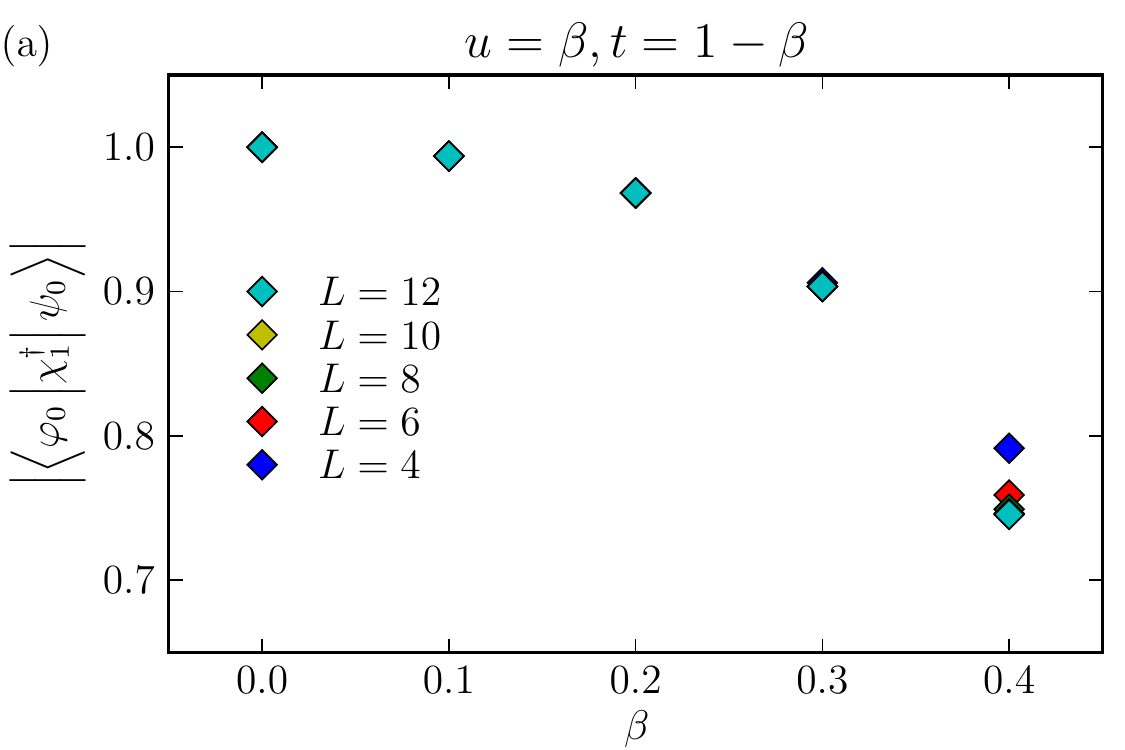}
\includegraphics[width=8cm]{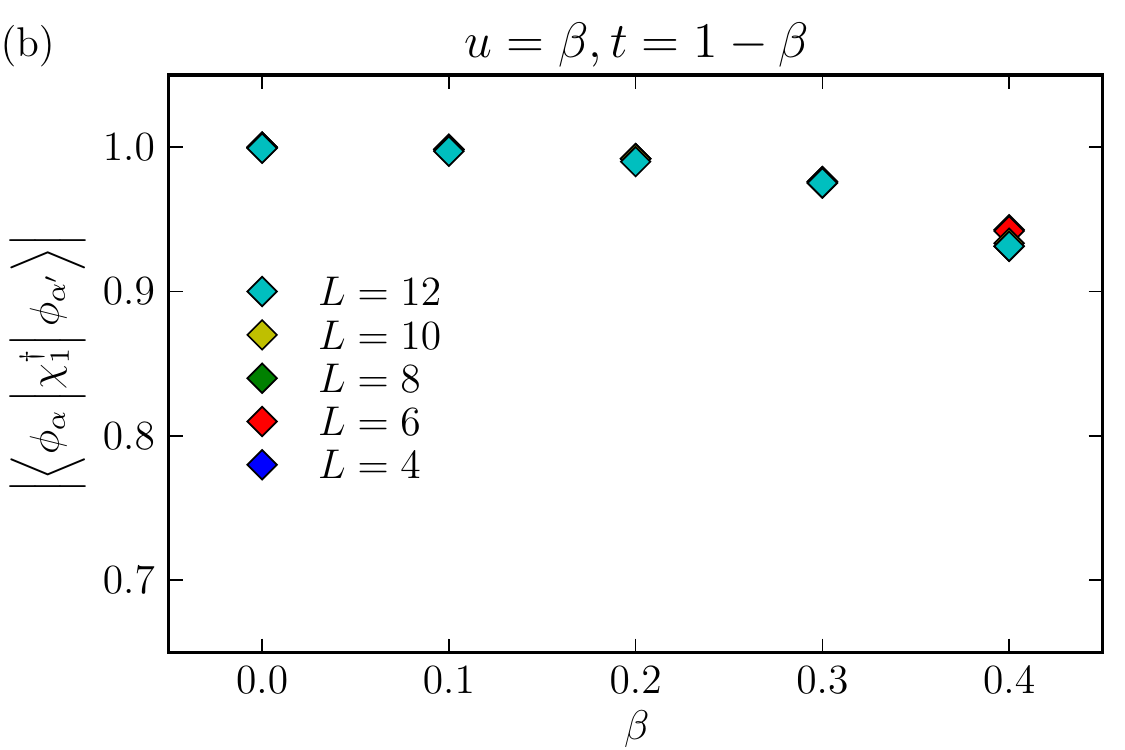}
\par\end{centering}
\caption{(a) Overlap of a ground state of the $\mathds{Z}_{4}$ parafermionic chain [Hamiltonian \eqref{eq:H_para} with $N=4$ for different $L$]
transformed by $\chi_{1}^{\dagger}$ and the ground state of the next $q$ sector. The state $\left| \psi_0 \right\rangle$
is the ground state for $q = 1$ and $\left| \varphi_0 \right\rangle$ for $q = -i$. (b) Matrix elements of $\chi^{\dagger}_{1}$ between two degenerate eigenstates $\left| \phi_{\alpha} \right\rangle$ and $\left| \phi_{\alpha^{\prime}} \right\rangle$ of the reduced density matrix $\rho^S$, which lie in  $q$-sectors that differ exactly by one charge. Calculations are performed for different chains of length $2L$ with periodic boundary conditions. The segment $S$ always includes the sites $1$ to $L$. \label{fig:transform}}
\end{figure}

The decomposition $Q_{\mathrm{eff}} \propto \chi_1^\dagger \chi_{2L}^{\vphantom{\dagger}}$ involving only operators of site 1 and site $2L$ is only exact in the limiting case
$u=0$, $t=1$. However, we expect $Q_{\mathrm{eff}}$ to be approximately given by a product of two operators, which are local at either ends of the chain even when $u\ne 0$, as long as one remains in the same phase. Each of these edge operators is localized over a region, which extends to a distance of the order of one correlation length from the boundary of the system. As long as the correlation length is finite, the exact edge operators at the left and the right ends of the chain have a finite overlap with $\chi_1$, $\chi_{2L}$.
To demonstrate this, we have diagonalized the Hamiltonian Eq.~(\ref{eq:H_Potts}) with open boundary conditions exactly for different chain lengths and various values of $\{u,t\}$, and calculated the matrix element of $\chi^{\dagger}_{1}$ between ground states with different eigenvalues of $Q$ . The results are shown in Fig.~\ref{fig:transform}(a). 
For ${t=1, u=0}$, the matrix element of $\chi_{1}^{\dagger}$ between ground states with neighboring eigenvalues of $Q$ is unity. For $u>0$, the matrix element decreases, but remains finite. The data clearly shows that the overlap does not depend on the system size but on the correlation length. As soon as the latter is comparable to the former ($\beta \approx 0.4$ for $L=4$), the matrix elements differ for larger system sizes.

Instead of looking at physical edge states of a  chain with open boundary conditions, we may also look at the ``virtual''  edge states we get by diagonalizing the reduced density matrix $\rho_S$. For this we consider again a bipartition of a periodic chain into a segment ($S$) of length $L_S$ and an environment ($E$) as discussed in Sec.~\ref{sec:fermion_class}. The reduced density matrix $\rho^S$ commutes with the charge operator $Q$, and thus the eigenstates $\left| \phi_{\alpha} \right\rangle$ of the segment have a defined charge $q$ modulo $N$. As in the case of fermions in Sec.~\ref{sec:fermion_class}, we claim that the charge operator acting on the important eigenstates fractionalizes into two operators acting on the two ends of the segment. These localized operators  then transform the important states with different  charge into one another. To support this claim, we calculated numerically the matrix elements of $\chi^{\dagger}_{1}$ between important eigenstates of $\rho_S$ with  $Q$-eigenvalues that 
differ by a charge of one parafermion. The results are shown in Fig.~\ref{fig:transform}(b).
%
Again, the overlap does not depend on the size of the segment as long as the correlation length is smaller than $L_S$. This strongly supports the fractionalization of $Q$ into a product of operators that are localized on either ends of the segment.


\subsection{Classifying phases by symmetry properties of the entanglement\label{subsec:class}}

We now discuss the classification of the possible phases of a
$\mathds{Z}_N$ parafermion chain, generalizing the technique used
to classify the phases of a fermionic chain
(Sec.~\ref{sec:fermion_class}). These phases are best understood
by first performing the Jordan-Wigner transformation
Eq.~(\ref{eq:mapping}), mapping the problem to a $\mathds{Z}_N$ spin
chain. In the Ising/Majorana chain (corresponding to $N=2$), the symmetry-broken phase of
the spin chain corresponds to a topologically nontrivial phase in
the fermionic variables, whereas the $\mathds{Z}_{2}$ symmetric
phase of the spin chain maps to the trivial phase.


In the $N$-state clock model, the $\mathds{Z}_{N}$ symmetry can
spontaneously break down to any subgroup of $\mathds{Z}_{N}$. The
number of subgroups is the number of divisors of $N$. If $N =
p_1^{r_1} \ldots  p_l^{r_l}$ is the prime factorization of $N$,
the number of divisors is $P = \prod_{i=1}^l (r_i+1)$. Therefore,
we expect to find $P$ distinct phases. It turns out that these
phases fall into two distinct classes. One class, analogous to
the nontrivial phase of the Majorana chain,  
will be referred to as \emph{symmetry fractionalized phases}. The other class can be described by the presence of a
non-zero order parameter, \empty{in the parafermionic
variables}, and will be referred to as \emph{parafermion condensate phases}. Both types phases are topological, in the sense that they lack a local (bosonic) order parameter. They can be distinguished by the ground-state degeneracy of a chain with open boundary conditions, as we discuss below. 



In order to find all the possible topological phases, let us
consider a chain of length $2L$ of $\mathds{Z}_{N}$ parafermions
with a Hamiltonian
$H\left(\chi_{1}^{\vphantom{\dagger}},\ldots,\chi_{2L}^{\vphantom{\dagger}}\right)$
that conserves the $\mathds{Z}_{N}$ charge $Q=\prod_{j=1}^L\left(
-e^{-\pi i/N} \chi_{2j-1}^{\dagger}\chi_{2j}^{\vphantom{\dagger}}
\right)$, hence
\begin{equation}
\left[H,Q\right]=0.
\end{equation}
We now consider a bipartition of the chain with periodic boundary
conditions into a segment (\ensuremath{S}) of length \ensuremath{L_S}
with $L_S\ll2L$ and an environment ($E$) as shown in Fig.~\ref{fig:segment}.

Just as in the case of fermions in Sec.~\ref{sec:fermion_class} and as supported by the numerical data in Fig.~\ref{fig:transform}(b),
we can write the operator $Q$ acting on the important states
$|\phi_{\alpha}\rangle$ as $Q\propto Q^{A}Q^{B}$, where $Q^{A}$ and
$Q^{B}$ are local operators acting on a finite region near the
left and right edges of $S$, respectively. The operator $Q$ itself is a
charge-neutral operator. However, $Q^{A}$ and $Q^{B}$ can carry
non-zero charges, such that $Q^{A} Q = e^{2\pi i k/N} Q Q^{A}$
where $k$ is an integer, and $Q^{B} Q = e^{-2\pi i k/N} Q Q^{B}$.
Different values of $k$ correspond to different topological
phases. Below, we derive a consistency condition which constrains
the allowed values of $k$.

We consider powers of the operators $\chi_j$ defined by
\begin{equation}
 \chi^{n_i}_j \equiv \chi^{N/p_i^{r_i}}_j, \quad i \in {1, \ldots,
 l},
\end{equation}
and the corresponding charge operators
\begin{equation}
 Q^{n_i} \equiv Q^{N/p_i^{r_i}}, \quad i \in {1, \ldots, l}.
\end{equation}
The original charge operator $Q$ can be written as a product
\begin{equation}
 Q = \left(Q^{n_1}\right)^{a_1} \ldots  \left(Q^{n_l}\right)^{a_l},\label{eq:Q_comb}
\end{equation}
with integer $a_1,\cdots,a_l$. We can again fractionalize $Q^{n_i}$
like the total charge operator as
\begin{equation}
 Q^{n_i} = \left(Q^{n_i}\right)^A \left(Q^{n_i}\right)^B.
\end{equation}
We now study the ways in which $Q^{n_i}$ can fractionalize. Similar to
above, while $Q^{n_i}$ is charge neutral, the operators $\left(Q^{n_i}\right)^A$,
$\left(Q^{n_i}\right)^B$ can carry charge $Q^{n_i}$. We may write $\left(Q^{n_i}\right)^A$ as
\begin{equation}
 (Q^{n_i})^A \propto \mathcal{B} \times  \left[\left(\chi^{n_i}_j\right)^{\dagger}\right]^{k_i}
\end{equation}
and correspondingly, $\left(Q^{n_i}\right)^B \propto \mathcal{B} \times
\left[\chi^{n_i}_j\right]^{k_i}$, where $\mathcal{B}$ denotes bosonic operators.

Commuting $\left(Q^{n_i}\right)^A$ and $\left(Q^{n_i}\right)^B$ leads to
\begin{equation}
 \left(Q^{n_i}\right)^A\left(Q^{n_i}\right)^B=e^{-\frac{2\pi i}{N}n_i^2 k_i^2}\left(Q^{n_i}\right)^B \left(Q^{n_i}\right)^A.\label{eq:k_ia}
\end{equation}
There is also another way to determine $k_i$.
Since $\left(Q^{n_i}\right)^A$ creates $n_i k_i$ parafermions, commuting $\left(Q^{n_i}\right)^A$ with $Q^{n_i}$ yields
\begin{eqnarray}
 \left(Q^{n_i}\right)^A Q^{n_i} = e^{-\frac{2\pi i}{N} n_i^2 k_i} Q^{n_i} \left(Q^{n_i}\right)^A
\end{eqnarray}
and therefore
\begin{eqnarray}
 \left(Q^{n_i}\right)^A \left(Q^{n_i}\right)^B = e^{-\frac{2\pi i}{N} n_i^2 k_i} \left(Q^{n_i}\right)^B \left(Q^{n_i}\right)^A.\label{eq:k_iq}
\end{eqnarray}
In order for \eqref{eq:k_ia} to be consistent with
\eqref{eq:k_iq}, we obtain a constraint on $k_i$:
\begin{equation}
\frac{n_i^2\left(k_i^2-k_i\right)}{N} = \frac{n_ik_i\left(k_i-1\right)}{p_i^{r_i}} = \text{integer}.\label{eq:cond}
\end{equation}
The only two solutions for $ p_i^{r_i} | k_i\left(k_i-1\right) $ are $k_i = 0$ or $k_i = 1$ since $k_i$ and $k_i-1$ cannot
be divisible by the same prime. Hence, each $Q^{n_i}$ can be fractionalized in two ways.

The two possibilities of fractionalizing the operators $Q^{n_i}$ together with Eq.~\eqref{eq:Q_comb} lead then to $2^l$ possible values of $k$ given by
\begin{equation}
 k = \sum_{i=1}^l a_i n_i k_i \ \text{mod }N, \label{eq:k}
\end{equation}
where each of the $k_i$ can be chosen as 0 or 1. Since there is no
number that is a common divisor of all summands in \eqref{eq:k},
each individual choice of ${k_i}$ corresponds to a different $k$.

We now derive the structure of the entanglement spectrum in the
different phases. The ground state wave function $|\psi_0\rangle$
has a well defined charge modulo $N$ and thus
$\left[\rho^{S},Q\right]=\left[\rho^{S},Q^{A}\right]=\left[\rho^{S},Q^{B}\right]=0$.
Hence, $Q$ and $\rho^{S}$ can be diagonalized simultaneously. For
$k \neq 0$, the eigenvalues of $\rho^{S}$ come in degenerate
multipletts which can be seen as follows.

If we have a state $|\phi_{\alpha}\rangle$
with $\rho^{S}|\phi_{\alpha}\rangle=\rho_{\alpha}^{S}|\phi_{\alpha}\rangle$
and $Q|\phi_{\alpha}\rangle=q_{\alpha}|\phi_{\alpha}\rangle$, then
$|\psi_{\alpha}\rangle=Q^{A}|\phi_{\alpha}\rangle$ has the eigenvalues
\begin{eqnarray}
\rho^{S}|\psi_{\alpha}\rangle & = & \rho_{\alpha}|\psi_{\alpha}\rangle,\\
Q|\psi_{\alpha}\rangle & = & e^{\frac{2\pi ik}{N}}q_{\alpha}|\psi_{\alpha}\rangle. \label{eq:ortho}
\end{eqnarray}
From \eqref{eq:ortho} we can see that $|\psi_{\alpha}\rangle$ and $|\phi_{\alpha}\rangle$
are orthogonal.

Every $\left(Q^{n_i}\right)^A$ can only change the charge of the state by $e^{\frac{2 \pi i}N n_ik_i } = e^{2 \pi i k_i/p_i^{r_i}}$.
Since different $p_i^{r_i}$ have no common divisor, we can act $D=\prod_{i=1}^l \left[ k_i \left( p_i^{r_i} -1 \right) + 1 \right]$
times with $Q^A$ on the state  $|\phi_{\alpha}\rangle$ until we obtain a state with the charge $q$ again. We identify this state as $|\phi_{\alpha}\rangle$.
The eigenvalues of $\rho_S$ are therefore at least $D$-fold degenerate.


We have identified $2^l$ different topological phases --- corresponding to phases in the clock model where the symmetry is broken
down to some subgroup of $\mathds{Z}_{N}$. The number of subgroups is the number of divisors of $N$, which is
$\prod_{i=1}^l (r_i+1)$. This number is greater than $2^l$ if any $r_i > 1$. What happened to the remaining phases in the parafermionic model?
Let us consider the example of a four-state clock model. Three phases can be realized, one with a ground state having the full $\mathds{Z}_{4}$ symmetry,
another $\mathds{Z}_{2}$-symmetric one and a completely symmetry broken one. The first and the third correspond to the trivial and the nontrivial
phase in the parafermionic chain, respectively. The second one is a 
 a parafermion condensate phase, characterized by a parafermionic order parameter. This can be seen as
follows. The squared parafermionic variables commute with each other,
\begin{equation}
 \left[\chi_i^2,\chi_j^2\right] = 0,
\end{equation}
thus they are mutually bosonic. 
Hence, we can have 
long-range order in the parafermionic variables:
\begin{equation}
 \left\langle \chi_i^2\chi_j^2 \right\rangle \neq 0.
\end{equation}
The argument that allows us to fractionalize the charge operator and identify the topological phase is no longer valid in the presence of symmetry breaking. Note that
a long range order in parafermionic variables is not possible if they obey a nontrivial exchange statistics. This is because in a system with a finite correlation length we have in the bulk
\begin{equation}
 \left\langle \chi_i\chi_{i+l} \right\rangle \rightarrow \left\langle \chi_i \right\rangle\left\langle\chi_{i+l} \right\rangle \quad \text{for }l \rightarrow \infty.
\end{equation}
If the operators do not commute, this expression has to go to zero. Otherwise it cannot be consistent with
\begin{equation}
 \left\langle \chi_i\chi_{i+l} \right\rangle = e^{2 \pi i/N} \left\langle \chi_{i+l}\chi_i \right\rangle .
\end{equation}

We now discuss the ground state degeneracies for systems with open and periodic boundary conditions in different phases. Consider, for example, a $\mathds{Z}_{4}$ chain. The ground state of the Hamiltonian 
\begin{equation}
 H_{\mathrm{SF}} = \sum_{j=1}^{L-1}\left[e^{\pi i/4}\chi_{2j}^{\dagger}\chi_{2j+1}^{\vphantom{\dagger}}+e^{-\pi i/4}\chi_{2j+1}^{\dagger}\chi_{2j}^{\vphantom{\dagger}}\right],
\end{equation}
is in the symmetry-fractionalized phase. This phase is characterized by four-fold ground state degeneracy for a chain with open boundary conditions, and a unique ground state for a chain with periodic boundary conditions. We have verified these expectations numerically. 

The ground state of the Hamiltonian 

\begin{eqnarray}
 H_{\text{PC}} &=& \sum_{j=1}^{L-1} \left[\left(\chi_{2j}^{\dagger}\right)^{2}\left(\chi_{2j+1}^{\vphantom{\dagger}}\right)^{2}+\left(\chi_{2j+1}^{\dagger}\right)^{2}\left(\chi_{2j}^{\vphantom{\dagger}}\right)^{2}\right]\nonumber \\
 &&+\sum_{j=1}^{L} \left[e^{\pi i/4}\chi_{2j-1}^{\dagger}\chi_{2j}^{\vphantom{\dagger}}+e^{-\pi i/4}\chi_{2j}^{\dagger}\chi_{2j-1}^{\vphantom{\dagger}}\right]
 \label{eq-sym-broken}
\end{eqnarray}
is in the parafermion condensate phase, in which the $\mathds{Z}_{4}$ symmetry is broken down to $\mathds{Z}_{2}$. The ground state of a chain with open boundary conditions is two-fold degenerate. In a chain with periodic boundary conditions, in which the term $\left(\chi^\dagger_{2L}\right)^2\left(\chi^{\vphantom{\dagger}}_{1}\right)^2 + \mathrm{H.c.}$ is added, the ground state of the Hamiltonian (\ref{eq-sym-broken}) is also two-fold degenerate. However, this ground state degeneracy is not stable; e.g., if we add the term $\chi^\dagger_{2L}\chi^{\vphantom{\dagger}}_{1}+\mathrm{H.c.}$, the two-fold degeneracy is lifted because $\chi_i$ does not commute with $\chi_j^2$. This demonstrates that the ground state degeneracy in the parafermion condensate phase is also topological in nature, and does not correspond to a local (bosonic) order parameter.


\subsection{Combination of chains}\label{sec:combine}

We now turn to investigate the topological phases that result when we combine parafermionic chains. As long as the interchain coupling is not too strong, the combined system is in a topological phase that is determined by the phases of the constituent chains. We derive an ``addition rule'' for chains in different phases. This is a generalization of the addition rules for phases of fermionic chains.\cite{Turner-2011}

\begin{figure}
\begin{centering}
\includegraphics[width=8cm]{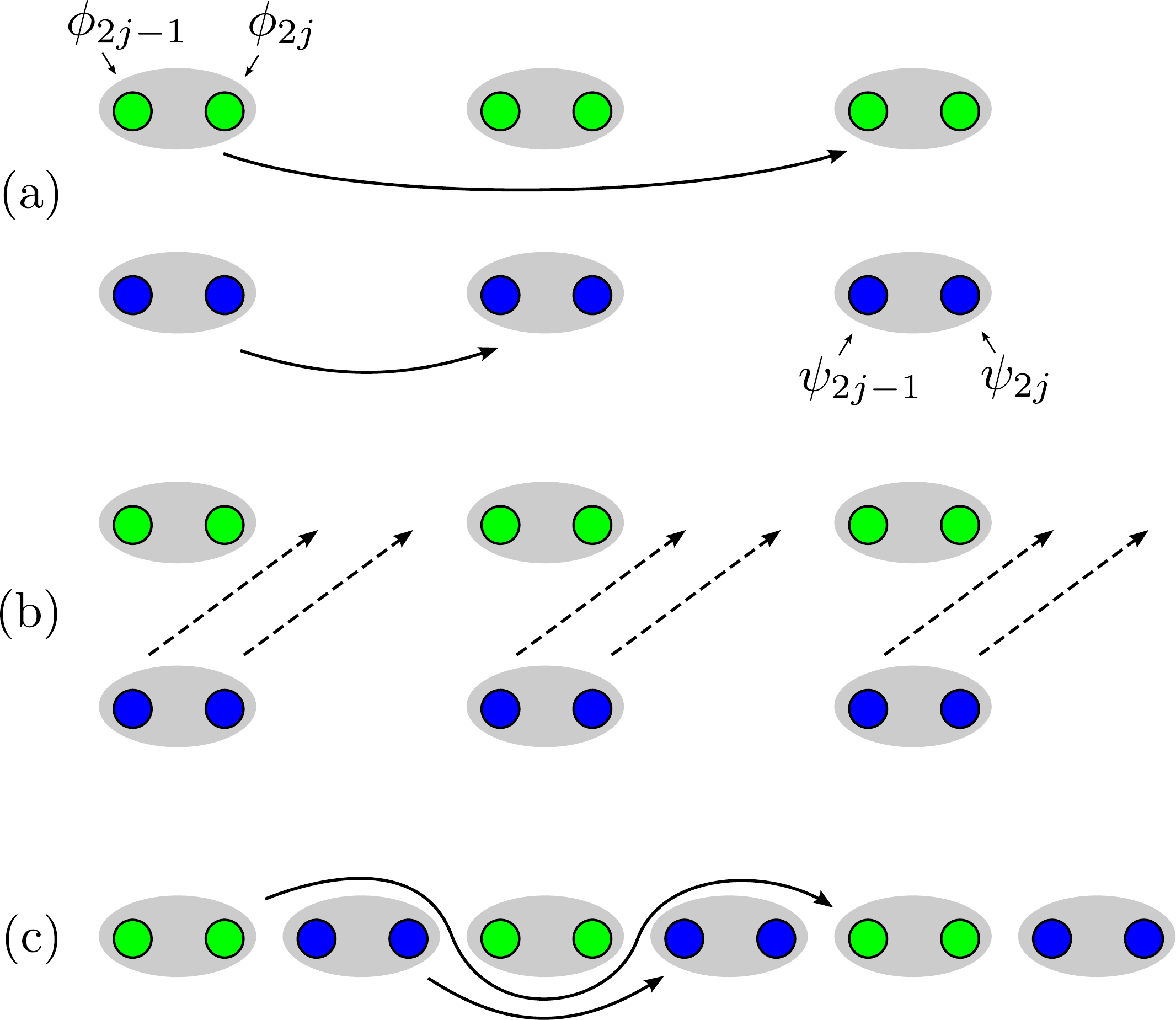}
\par\end{centering}

\caption{Combination of two chains. The hopping in the two independent chains is shown in (a). After combining the chains (b) not all hopping processes occur below the chain. The anyons of former
chain $\phi$ pass above the sites of former chain $\psi$ (c). \label{fig:combine}}
\end{figure}

Let us consider two chains with parafermionic operators $\phi_{2j-1}, \phi_{2j}$
for the first one and $\psi_{2j-1}, \psi_{2j}$ for the second. The setup is shown schematically in Fig.~\ref{fig:combine}(a). Our first task is to combine the two chains into a single effective chain with parafermion operators  $\chi_{4j-3},\chi_{4j-2},\chi_{4j-1}, \chi_{4j}$, see Fig.~\ref{fig:combine}(b). 

Suppose that the Hamiltonians of the two chains, $H_\psi$ and $H_\phi$, are given . In order to construct the Hamiltonian of the combined chain, $H_\chi$, we imagine the two chains embedded in a two-dimensional plane. The parafermions are thought of as Abelian anyons which acquire a phase of $e^{2\pi i/N}$ when exchanged counterclockwise. We must then pay attention to the paths the anyons hop along. Hopping processes that go above or below the chain correspond, in general, to \emph{different} operators. E.g., motion above the chain, and then motion back below the chain, corresponds to an anyon moving in a closed circle, and should therefore give a phase that depends on the total anyonic charge enclosed by the path. 


We define the operator $\phi^\dagger_i \phi^{\vphantom{\dagger}}_j$ to describe hopping between sites $i$ and $j$ in the first chain, such that the anyons move \emph{below} the chain. Similarly, $\psi^\dagger_i \psi^{\vphantom{\dagger}}_j$ and $\chi^\dagger_i \chi^{\vphantom{\dagger}}_j$ describe hopping processes between sites in the $\psi$ and $\chi$ chains that occur below the chains. After combining the two chains, however, we must consider the fact that the hopping process between $\phi$ sites follows a trajectory that  passes the $\psi$ sites from~\emph{above} [see Fig.~\ref{fig:combine}(c)]. To take that into account, we define the following rule to map operators acting on the $\phi$ chain to those of the $\chi$ chain:


\begin{eqnarray}
 \phi^{\dagger}_{2i-1} \phi^{\vphantom{\dagger}}_{2j-1} &\rightarrow& \chi^{\dagger}_{4i-3} \chi^{\vphantom{\dagger}}_{4j-3} \label{eq:convert} \\ \nonumber
 &&\times \left(\prod_{i\leq k < j} e^{\pi i/N} \chi_{4k-1}^{\vphantom{\dagger}} \chi_{4k}^{\dagger} \right)^{2},
\end{eqnarray}
where have assumed that $i<j$, and that the hopping occurs between $\phi$ sites with odd indices. In this equation, we may replace $\{\phi^\dagger_{2i-1}, \chi^{\dagger}_{4i-3} \}$ by $\{\phi^\dagger_{2i}, \chi^{\dagger}_{4i-2} \}$, respectively, corresponding to a hopping process originating from a $\phi$ site with an even index, and similarly $\{\phi_{2j-1}, \chi_{4j-3} \}$ can be replaced by $\{\phi_{2j}, \chi_{4j-2} \}$.

The additional factor in brackets at the end of Eq.~\eqref{eq:convert} emerges as follows. 
The operator $\chi^\dagger_{4i-3} \chi^{\vphantom{\dagger}}_{4j-3}$ describes a process in which a particle hops below the combined chain. The fact that hopping between former $\phi$ sites passes the $\psi$ sites from above is implemented by next hopping back below each $\psi$ site which was passed, and then hopping forward again, this time above the $\psi$ site. This operation leads to a phase factor of $e^{-4 \pi i/N}$ to the power of the number of parafermions at the $\psi$ site. This is exactly the factor in brackets in Eq.~\eqref{eq:convert}. Note that this issue does not arise in Majorana chains ($N=2$) where a phase of either $e^{+i\pi}$ or $e^{-i\pi}$ appears when hopping below or above the chain, respectively, so the additional phase factor always squares to unity.

Following the same logic, hopping processes within the $\psi$ chain do not need to be modified when mapping them to the combined chain:

\begin{equation}
\psi^{\dagger}_{2i-1} \psi^{\vphantom{\dagger}}_{2j-1} \rightarrow \chi^{\dagger}_{4i-1} \chi^{\vphantom{\dagger}}_{4j-1}.
\end{equation}
A similar rule holds for $\psi$ sites with even indices.

Let us now turn to study the addition rule of topological phases in the $\phi$ and $\psi$ chains. The fractionalization rule for the charge operator of the $\phi$ chain, $Q(\phi)$,  can be affected by the considerations above. We need to map $Q(\phi)$ to the language of the combined chain $\chi$ operators according to the rule of Eq.~\eqref{eq:convert}. If we name the $\chi$ operators on the former $\phi$ sites as $\phi'$
and the ones on the former $\psi$ sites $\psi'$, we obtain
\begin{equation}
 Q^A (\phi) Q^B(\phi) \rightarrow Q^A(\phi') Q^B (\phi') Q(\psi')^{-2k_{\phi}}, \label{eq:string}
\end{equation}
where the original $Q^{A,B}(\phi)$ operators are fractionalized according to $Q^{A,B}(\phi)\sim \phi^{\pm k_\phi}$, respectively. If we combine the two chains, we can calculate the $k$ characterizing the fractionalization of the charge operator in the combined chain as follows:
\begin{eqnarray}
 Q_{\text{total}} &=& Q(\phi) Q(\psi) \nonumber \\
 &=& Q (\phi') Q(\psi')^{-2k_{\phi}} Q (\psi') = Q (\phi') Q(\psi')^{1-2k_{\phi}} \nonumber \\
 &\propto& \underbrace{Q^A (\phi') Q^A(\psi')^{1-2k_{\phi}} }_{Q_{\text{total}}^A}\underbrace{Q^B (\phi') Q^B(\psi')^{1-2k_{\phi}}}_{Q_{\text{total}}^B}\nonumber \\
\end{eqnarray}
In the last expression we omitted an overall phase factor that we get from commuting the charge operators. We obtain for the resulting state 
\begin{equation}
 k_{\text{total}} = k_{\phi} + k_{\psi} - 2 \,k_{\phi}k_{\psi} \label{eq:combine_k}
\end{equation}
The new $k_{\text{total}}$ fulfills the consistency condition Eq.~\eqref{eq:cond} since
\begin{equation}
 k_{\text{total}}^2 - k_{\text{total}} = k_{\phi}^2 - k_{\phi} + k_{\psi}^2 - k_{\psi} + 4\left( k_{\phi}^2 - k_{\phi} \right) \left( k_{\psi}^2 - k_{\psi} \right),
\end{equation}
and each $k^2_{\phi/\psi}-k_{\phi/\psi}$ separately is a multiple of $N$.
If $N$ is a prime number, then there exist only two phases $k=0,1$. Combining two chains with $k_{\phi}=k_{\psi}=1$, we find that the resulting phase is characterized by $k_{\text{total}}=0$. Thus, the addition rule of phases has a $\mathbb{Z}_2$ group structure.

\subsection{Presence of additional symmetries}
When there are extra symmetries in addition to the fundamental $\mathbb{Z}_N$ symmetry related to charge conservation, a richer classification of phases arises. Interestingly, as we will demonstrate below, the phases of parafermion chains can have non-commutative group structures. This is in contrast to the symmetry-protected phases of fermions or bosons in one dimension which are always described by commutative groups, even
if the physical symmetry group is non-commutative.\cite{ChenGu-2011, Pollmann-2012b}  

Consider, for example, $\mathbb{Z}_N$ parafermions, where $N$ is prime, with an extra $\mathbb{Z}_N$ symmetry $R$ that describes an internal degree of freedom (e.g., an orbital degrees of freedom). The phases of such a system are classified in terms of how $Q$ and $R$ fractionalize into terms that act on the edges of a segment: besides the parafermion charge of $Q_A$, $k$, we also have to keep track of $l$, the parafermion charge of $R_A$. Altogether
there are $2N$ possible phases, since $k$ is constrained to be $0$ or $1$ (see Sec.~\ref{subsec:class}), while $l$ can be any integer modulo $N$.
Note that the second integer is similar to the one that classifies symmetry protected topological phases of bosons with $\mathbb{Z}_N \times \mathbb{Z}_N$ symmetry\cite{Pollmann-2010, ChenGu-2011}: If there are
two symmetries $R$, $S$, the charge of the fractionalized operators $R_A$ under the action of $S_A$ can be expressed
as $S_A^{\vphantom{-1}}R_A^{\vphantom{-1}}S_A^{-1}R_A^{-1}=e^{{2\pi i l}/{N}}$ where $l$ classifies the different phases.

All the possible phases can be formed  by combining copies of two systems, one in a phase that we label by $X$ and the other in a phase labeled by $Y$. The $X$ phase is of order $2$, meaning that combining two systems in the $X$ phase results in a system in the trivial phase. The $Y$ phase is of order $N$. The phase $X$ is the parafermion nontrivial phase we studied before, where all the particles
are assumed to have zero charge under $R$ (i.e., the operators $R_{A/B}$ into which $R$ fractionalizes do not carry any charge); it is defined by $k=1$ and $l=0$.  The other phase, $Y$, is defined by $k=0$ and $l=1$, and it contains particles that have a charge under $R$ as well as $Q$.  We use the notation  $(k,l)$ to represent the combination of phases $X^kY^l$. 

These phases
follow the same multiplication rules as the dihedral group (the symmetries of an $N$-sided polygon where $X$ corresponds to a reflection and $Y$ corresponds to a rotation). To show this,
we must find how to classify a product of two states $|\phi\rangle\otimes|\psi\rangle$.
Following the arguments of Sec.~\ref{sec:combine}, when the two chains are combined, $R(\phi)$ gets a string added onto it, given
by a power of $Q(\psi)$: $R\rightarrow R^A(\phi')R^B(\phi')Q(\psi')^{-2l_\phi}$ [see Eq.~\eqref{eq:string}; recall that $R^A(\phi)$ and $R^B ({\phi})$ carry a charge of $\pm l_\phi$, respectively].  Thus,
\begin{eqnarray}
R_{\text{total}} &\propto& R^A(\phi')R^A(\psi')Q^{A}(\psi')^{-2l_\phi} \nonumber \\
&&\times R^B(\phi')R^B(\psi')Q^{B}(\psi')^{-2l_\phi}
\end{eqnarray}
so
\begin{equation}
l_{\text{total}}=l_\phi(1-2k_\psi)+l_\psi.
\end{equation}
Together with the addition rules for $k$ from Eq.~(\ref{eq:combine_k}) we find that  
\begin{equation}
(k_\phi,l_\phi)\times (k_\psi,l_\psi)=(k_\phi+k_\psi-2k_\phi k_\psi,l_\phi(1-2k_\psi)+l_\psi).
\end{equation}

In particular, we find for any prime number $N$ that $Y \times X= X\times Y^{-1}$: the phases do not commute for $N>2$.  This shows
that the group is the dihedral group (which is non-Abelian for $N>2$).

As a concrete example,  we obtain for the case of $N=3$ the following multiplication table:
\begin{center}
\begin{tabular}{c|cccccc}
$\psi / \phi $&(0,0)&(0,1)&(0,2)&(1,0)&(1,1)&(1,2)\\
\hline
(0,0)&(0,0)&(0,1)&(0,2)&(1,0)&(1,1)&(1,2)\\
(0,1)&(0,1)&(0,2)&(0,0)&(1,1)&(1,2)&(1,0)\\
(0,2)&(0,2)&(0,0)&(0,1)&(1,2)&(1,0)&(1,1)\\
(1,0)&(1,0)&(1,2)&(1,1)&(0,0)&(0,2)&(0,1)\\
(1,1)&(1,1)&(1,0)&(1,2)&(0,1)&(0,0)&(0,2)\\
(1,2)&(1,2)&(1,1)&(1,0)&(0,2)&(0,1)&(0,0)\\
\end{tabular}
\end{center}
\label{default}
This is exactly the multiplication table for the dihedral group of order 6.

\section{The physical realization of different topological phases\label{sec:physical_system}}

We begin by briefly reviewing a physical setup, which gives rise to parafermionic quasiparticles at the edge of a fractional topological insulator.
We then extend the setup by coupling of the localized quasiparticles through different media, which allows to tune the system into different topological phases.

\subsection{Parafermions on the edge of a fractional topological insulator}

\begin{figure}
\begin{centering}
\includegraphics[width=7cm]{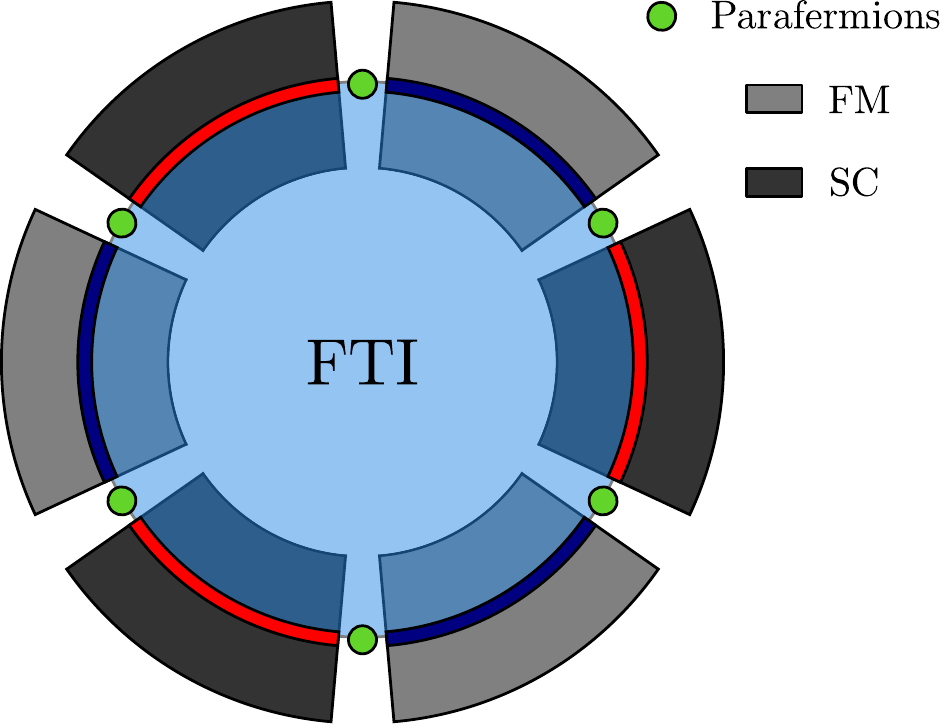}
\par\end{centering}

\caption{Physical setup: The edge modes of
the two-dimensional fractional topological insulator are gapped out by coupling to
superconductiong and ferromagnetic domains. At the domain walls, localized
parafermionic modes emerge. \label{fig:fti_droplet}}
\end{figure}

We consider the edge of a two-dimensional fractional topological
insulator (FTI) \cite{LevinStern2009}, which can be thought of as
a stack of two FQH states, with filling fraction
$\nu_{\uparrow}=1/m$ for spin up electrons and
$\nu_{\downarrow}=-1/m$ for spin down electrons, respectively. On
the edge, there are two counterpropagating edge modes. In analogy
with the edge of a quantum spin Hall insulator \cite{XuFu2010},
these edge modes can be gapped by proximity to either a
superconductor or to a ferromagnet\cite{Lindner,Cheng}. The same
physics can arise by bringing two counter-propagating FQH edge
states close together and coupling them either directly or by
proximity to a superconductor.\cite{Clarke,Lindner} For
concreteness, let us consider an FTI disk with an array of
alternating $L$ superconducting and $L$ ferromagnetic domains on
its perimeter (see Fig.~\ref{fig:fti_droplet}).

The low-energy theory of the edge in these setups has been shown
to be closely related to the parafermionic chain.\cite{Clarke} At
each interface between a ferromagnetic and a superconducting
region, there is a localized zero mode. Zero modes of different
interfaces satisfy anyonic commutation relations, equivalent
to those of Eq. (\ref{eq:comm}) with $N=2m$. It is therefore
natural to identify each zero mode with a parafermionic operator.
If two interfaces are sufficiently close to each other, fractional
quasiparticles can tunnel between them. The resulting low-energy
Hamiltonian is of the form of Eq. (\ref{eq:H_para}), where the
parafermion hopping terms describe quasiparticle tunneling
processes between different interfaces.

Below, we demonstrate for the example of $N=6$ that an
appropriately designed FTI system can realize all the phases of a
$\mathbb{Z}_{N}$ parafermionic chain.


\subsection{Filling fraction \ensuremath{\nu=1/3} \label{sec:filling-fraction_1o3}}

\begin{figure}[b]
\begin{centering}
\includegraphics[width=7cm]{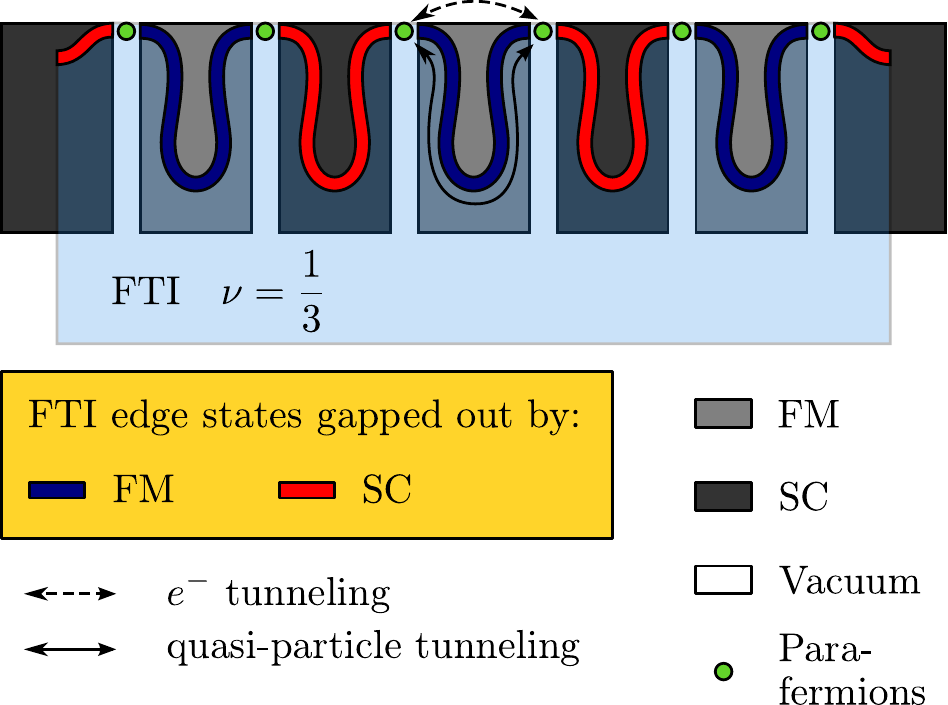}
\par\end{centering}

\caption{Physical realization of the Hamiltonian $H_{0}$ \eqref{eq:H_0}
 \label{fig:fti_edge} on the edge of
a fractional topological insulator (FTI). Changing the geometry can
be used to tune the effective tunneling amplitudes for different particles.}
\end{figure}

Let us consider the $\mathbb{Z}_{6}$-charge-conserving model, which
arises at the boundary of an FTI with filling factor 1/3. We now
discuss the physical realization sketched in
Fig.~\ref{fig:fti_edge}, which allows us to access four different
topological phases.

The edge states of a $\nu=1/3$ topological insulator are
alternatingly coupled to superconducting and ferromagnetic regions
as explained in the previous subsection. If the regions are short
enough, fractional quasiparticles can tunnel between adjacent
SC-FM interfaces. However, since the interfaces are only separated
by a small distance of vacuum, electrons can tunnel between them
as well. In terms of the parafermionic degrees of freedom this
leads to a term coupling triples of parafermions on adjacent
sites. By varying the length of the regions and the geometry of
the edge, the amplitudes of the two types of tunneling processes
can be tuned relative to each other.
Putting the above tunneling terms together, we obtain the following effective  Hamiltonian:
\begin{eqnarray}
H_{0} = \sum_{j=1}^{L-1} &&\left\{-i t_3\left[\left(\chi_{2j}^{\dagger}\right)^{3}\left(\chi_{2j+1}^{\vphantom{\dagger}}\right)^{3}-\left(\chi_{2j+1}^{\dagger}\right)^{3}\left(\chi_{2j}^{\vphantom{\dagger}}\right)^{3}\right]\right.\nonumber \\
 &  & \left.+ \, t_1 \left[ e^{\pi i/6} \chi_{2j}^{\dagger}\chi_{2j+1}^{\vphantom{\dagger}}+e^{-\pi i/6} \chi_{2j+1}^{\dagger}\chi_{2j}^{\vphantom{\dagger}}\right]\right\}\nonumber \\
+\sum_{j=1}^{L}  &  & \left\{ u_1\left[e^{\pi i/6}\chi_{2j-1}^{\dagger}\chi_{2j}^{\vphantom{\dagger}}+e^{-\pi i/6}\chi_{2j}^{\dagger}\chi_{2j-1}^{\vphantom{\dagger}}\right]\right. \label{eq:H_0} \\
  &  & \left.-\, iu_3\left[\left(\chi_{2j-1}^{\dagger}\right)^{3}\left(\chi_{2j}^{\vphantom{\dagger}}\right)^{3}-\left(\chi_{2j}^{\dagger}\right)^{3}\left(\chi_{2j-1}^{\vphantom{\dagger}}\right)^{3}\right]\right\} \nonumber,
\end{eqnarray}
which describes the hopping of single parafermions or three
parafermions at once. The operators \ensuremath{\chi_i^{3}} obey the
algebra of Majorana fermions:
\begin{equation}
\left\{ \chi_{i}^{3},\chi_{j}^{3}\right\} = 2\,\delta_{ij}
\end{equation}
From the classification described in the previous section, we know that there are four distinct phases in a chain of $\mathds{Z}_{6}$ parafermions. Each phase is characterized by a specific degeneracy of the entanglement energy levels. We now show that all four phases can be realized by tuning the parameters in the Hamiltonian $H_0$ \eqref{eq:H_0}. A chain of $\mathds{Z}_{6}$ parafermions may be viewed as a combination of a $\mathds{Z}_{3}$ parafermionic chain and a $\mathds{Z}_{2}$ (Majorana) chain. The corresponding conserved charges of these chains are given by the operators $Q^{n_i}$ introduced in Sec.~\ref{subsec:class}. The operator $Q^3$ is the conserved charge of the Majorana chain and $Q^2$ is the charge of the $\mathds{Z}_{3}$ chain. 

Let us introduce the operators
\begin{eqnarray}
 \eta_i &=& \chi_i^2, \\
 \gamma_i &=& \chi_i^3,
\end{eqnarray}
having the properties
\begin{eqnarray}
 \eta_i^3 &=& 1, \\
 \eta_i \eta_j &=& e^{-2\pi i / 3} \eta_j \eta_i, \quad \text{for } i < j, \\
 \gamma_i^2 &=& 1, \\
 \left\{ \gamma_i, \gamma_j \right\} &=& 2 \delta_{ij}, \\
 \left[ \eta_i, \gamma_j \right] &=& 0.
\end{eqnarray}
$H_0$ may then be written as
\begin{eqnarray}
 H_0 &=& \sum_j \left[ t_1 \left( e^{\pi i/6} \gamma_{2j}^{\vphantom{\dagger}} \gamma_{2j+1}^{\vphantom{\dagger}} \eta_{2j}^{\dagger} \eta_{2j+1}^{\vphantom{\dagger}} + \text{H.c.} \right) \right. \nonumber \\
 && + u_1\left(e^{\pi i/6} \gamma_{2j-1}^{\vphantom{\dagger}} \gamma_{2j}^{\vphantom{\dagger}} \eta_{2j-1}^{\dagger} \eta_{2j}^{\vphantom{\dagger}} + \text{H.c.} \right) \nonumber \\
 && \left. - 2i \,\left( t_3 \, \gamma_{2j}^{\vphantom{\dagger}} \gamma_{2j+1}^{\vphantom{\dagger}} + u_3 \, \gamma_{2j-1}^{\vphantom{\dagger}} \gamma_{2j}^{\vphantom{\dagger}} \right)\right].
\end{eqnarray}
To analyze this Hamiltonian, let us employ a mean field approximation in which we set
\begin{eqnarray}
 \gamma_{2j}^{\vphantom{\dagger}} \gamma_{2j+1}^{\vphantom{\dagger}} \eta_{2j}^{\dagger} \eta_{2j+1}^{\vphantom{\dagger}}&& \\
 \rightarrow &-&\frac i2 \underbrace{\left\langle i\gamma_{2j}^{\vphantom{\dagger}} \gamma_{2j+1}^{\vphantom{\dagger}} \right\rangle}_{\Gamma_{2j}} \eta_{2j}^{\dagger} \eta_{2j+1}^{\vphantom{\dagger}} \nonumber \\ 
 &+& \frac 12 e^{\pi i/3} \gamma_{2j}^{\vphantom{\dagger}} \gamma_{2j+1}^{\vphantom{\dagger}} \underbrace{ \left\langle e^{-\pi i/3} \eta_{2j}^{\dagger} \eta_{2j+1}^{\vphantom{\dagger}} \right\rangle}_{\Delta_{2j}}. \nonumber
\end{eqnarray}
In this approximation we obtain an effective mean-field Hamiltonian $H_{\text{MF}} = H_\gamma + H_\eta$ with
\begin{eqnarray}
 H_\gamma &=& - \sum_j \left[ i\left(t_3- \frac{t_1}{2} \Delta_{2j} \right) \gamma_{2j}^{\vphantom{\dagger}} \gamma_{2j+1}^{\vphantom{\dagger}} \right. \nonumber \\
  && \left. +i \left(u_3- \frac{u_1}{2} \Delta_{2j-1} \right) \gamma_{2j-1}^{\vphantom{\dagger}} \gamma_{2j}^{\vphantom{\dagger}}  + \text{H.c.} \right],
\end{eqnarray}
and
\begin{eqnarray}
 H_\eta &=& \sum_j \left[ \frac{t_1}2 \Gamma_{2j} e^{-\pi i/3} \eta_{2j}^{\dagger} \eta_{2j+1}^{\vphantom{\dagger}}  \right. \nonumber \\
  && + \left. \frac{u_1}2 \Gamma_{2j-1} e^{-\pi i/3} \eta_{2j-1}^{\dagger} \eta_{2j}^{\vphantom{\dagger}} + \text{h.c.} \right].
\end{eqnarray}
Hence, this effective model leads to two decoupled chains, a Majorana and a $\mathds{Z}_{3}$ parafermion chain. Of course, this approximation is rather crude. However, we can use it to estimate in which parameter region the four phases occur.

\begin{figure}[t]
\begin{centering}
\includegraphics[width=8cm]{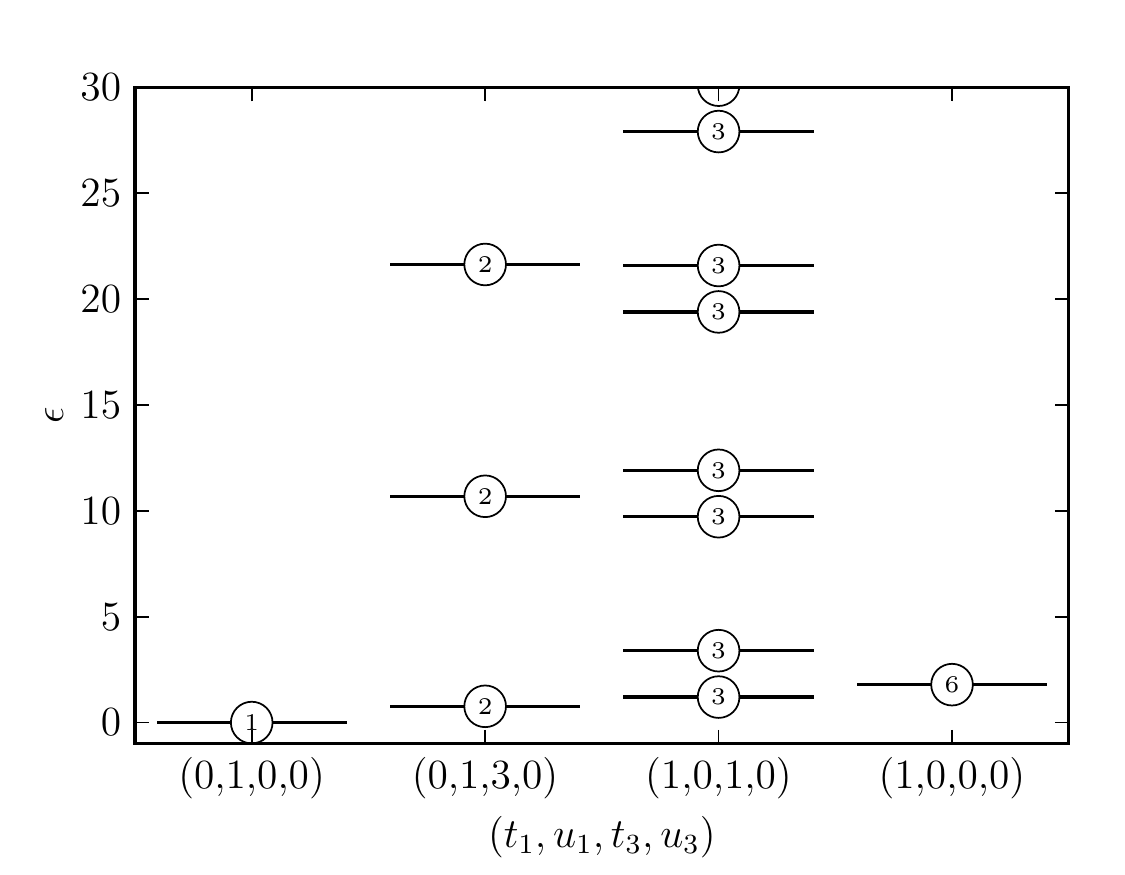}
\par\end{centering}

\caption{Entanglement spectra for different values  of the parameters $( t_1,u_1,t_3,u_3)$ in $H_0$. 
\label{fig:entanglement}}
\end{figure}

To prove the existence of the four phases for the Hamiltonian $H_0$ \eqref{eq:H_0}, we investigate the entanglement spectra for different parameter values numerically by exact diagonalization of chains of length $L=6$ in the clock model variables corresponding to 12 superconducting and ferromagnetic domains, with open boundary conditions. The results are shown in Fig.~\ref{fig:entanglement}.

The expected degeneracies in the entanglement spectrum are clearly reproduced. The ground states for $\left( t_1, u_1,t_3,u_3 \right) = \left( 0,1,0,0 \right)$ and $\left( 1,0,0,0 \right)$ are fully dimerized states
with the pairing of parafermions either within the same sites or between neighboring sites. The entanglement spectrum in these phases consists of a single multiplet of degeneracy 1 and 6, respectively. For $\left( t_1, u_1,t_3,u_3 \right) = \left( 0,1,3,0 \right)$ we get that each level in the entanglement spectrum is two-fold degenerate, corresponding to the nontrivial phase of the $\gamma$ (Majorana) sector\cite{Fidkowski-2010a}, whereas the $\eta$ sector is in the trivial phase. Similarly, for $\left( t_1, u_1,t_3,u_3 \right) = \left( 0,1,0,1 \right)$, each entanglement level is three-fold degenerate, corresponding to the nontrivial phase of the $\eta$ sector. We have therefore shown that all of the phases predicted by our classification scheme are realized in the proposed model.

\section{Conclusions}\label{sec:Conclusions}
In this paper we derived a systematic procedure for classifying the phases of  a 1D parafermionic system. The procedure is an extension of the previously introduced classification scheme of 1D interacting fermions. We have shown that a 1D parafermionic system supports symmetry fractionalized as well as parafermion condensate phases. The symmetry fractionalized phases are characterized by a fractionalization of the operator, which measures the conserved charge on a segment, while the parafermion condensate phases are identified by long-range correlations of an order parameter composed of multiple parafermion operators. Different topological phases can be identified by  degeneracies in the entanglement spectrum. 

We have derived an addition rule for chains in different phases. We furthermore showed that in the presence of additional symmetries the phases of parafermions can form non-commutative groups.  As a concrete physical realization of the model we studied the $\nu=1/3$ fractional topological insulator, which has been shown to yield a low-energy effective $N=6$ parafermionic edge model. Here we showed that a simple physical setup with two both electron and quasiparticle tunneling allows us to realize all four possible topological phases.

\emph{Note added in proof.} Recently, a closely related work appaered with similar conclusions.\cite{Quella2013}

\acknowledgements{We thank Netanel Lindner for useful discussions. E. B. was supported by the ISF under Grant No. 7113640101, and by the Robert Rees Fund.}


\end{document}